\documentclass[a4paper, 11pt]{cas-sc}

\usepackage[square,numbers]{natbib}

\usepackage{graphicx}
 \usepackage{amsmath} 
 \usepackage{subcaption}
 \usepackage{multirow}
 \usepackage{url}
\usepackage{algorithm}
\usepackage{algorithmicx}
\usepackage{algpseudocode}

\def\tsc#1{\csdef{#1}{\textsc{\lowercase{#1}}\xspace}}
\tsc{WGM}
\tsc{QE}

\def\nchannel{p}
\def\efs{f_s}
\def\eeg{e}
\def\speech{s}
\def\sfs{F_s}
\def\feeg{E}
\def\seeg{\speech}
\def\fseeg{S}
\def\f{k}
\def\F{\f'}
\def\t{n}
\def\nofsamples{N}
\def\tspeech{\t'}
\def\T{\nofsamples}
\def\sfeeg{E^{stretch}}

\def\ieeedata{{\sc eegdot}}
\def\EEG{{\sc eeg}}
\def\participants{Subjects}

\begin{document}
\let\WriteBookmarks\relax
\def\floatpagepagefraction{1}
\def\textpagefraction{.001}

\shorttitle{Signal Transformation}    

\shortauthors{Sunil Kopparapu}  

\title [mode = title]
{Signal Transformation for Effective Multi-Channel Signal Processing}


%

\author[1]{Sunil Kumar Kopparapu}
[orcid=0000-0002-0502527X,linkedin=sunilkopparapu]

\cormark[1]


\ead{sunilkumar.kopparapu@tcs.com}

\ead[url]{http://www.tcs.com}

\credit{Conceptualization of this study, Methodology, Software}

\affiliation[1]{organization={TCS Research},
            addressline={Tata Consultancy Services Limited}, 
            city={Mumbai},
            postcode={400601}, 
            state={Maharastra},
            country={India}}







\cortext[1]{Corresponding author}



\begin{abstract}
Electroencephalography (EEG) is an non-invasive method to record the electrical activity of the brain. The EEG signals are low bandwidth and recorded from multiple electrodes simultaneously in a time synchronized manner. Typical EEG signal processing involves extracting features from all the individual channels separately and then {\em fusing} these features for downstream applications. In this paper, we propose a signal transformation, using basic signal processing, to combine the individual channels of a low-bandwidth signal, like the EEG into a single-channel high-bandwidth signal, like audio. Further this signal transformation is bi-directional, namely the high-bandwidth single-channel can be transformed to generate the individual low-bandwidth signals without any loss of information. Such a transformation when applied to EEG signals overcomes the need to process multiple signals and allows for a single-channel processing. 
The advantage of this signal transformation is that it allows the use of pre-trained single-channel pre-trained models, for multi-channel signal processing and analysis. We further show the utility of the signal transformation on publicly available EEG dataset. 

\end{abstract}




\maketitle


\section{Introduction}
\label{sec:introduction}

Electroencephalography (EEG) is a method to record the spatial electrical activity
of the brain non-intrusively. Typically the instrument used to record EEG has several electrodes mounted on a cap which when placed on the head can sense the electrical activity of the brain at the location or in the spatial vicinity of the electrode. Each electrode captures the brain activity underneath the electrode at a low sampling rate, typically between $250$ and $1000$ Hz. As a consequence if one uses an EEG instrument with $n$ electrodes, there are $n$-channels of signals that are simultaneously recorded from each of electrode and are time synchronized. So any analysis on the EEG signal implies processing all the $n$ electrode output signals together to exploit the property that these signal are time synchronized. This, EEG signal, can be considered as an example of a multi-channel signal.

There are several disadvantages and challenges when processing multi-channel signals, like EEG. 
%
For example, to mention a few,
(a) multi-channel  data involves multiple signals recorded simultaneously, which 
makes it more difficult to identify relevant patterns and interpret results compared to single-channel signals, (b)
processing and analyzing multi-channel data requires more computational resources and time
longer processing times, 
(c)
different channels may be correlated due to shared sources of brain activity or artifacts, throwing up challenges during analysis, 
use of 
techniques to disentangle the signals, (d)
interpretation of results from multi-channel EEG can be difficult, relating the observed activity in different channels to specific cognitive processes or brain regions, (e)
lack of standardization in terms of the number of electrodes used to record EEG data or the sampling rate at which the signal is digitized 
leads to variability in results across studies; making it difficult to compare findings or even build pre-trained foundational models as is common in single-channel signals like audio and speech, (f)
a risk of overfitting models to the data, especially if the number of data samples is not sufficiently large relative to the number of channels, leading to poor generalization while analyzing unseen data, and
(g) visualizing multi-channel EEG data requires methods to represent both the spatial and temporal aspects in a way that is interpretable.
In this paper, we propose a simple yet effective \textit{signal transformation} that overcomes some of these disadvantages of processing and analyzing a multi-channel signal. The proposed transformation allows for representation of a multi-channel signal as a single-channel signal where the transformed signal retains the properties of the original multi-channel signal, making the transformation reversible. The single-channel signal not only allows for better visualization of the signal, but also allows for use of pre-trained foundational models that are available in plenty for single-channel signals, like speech and audio.


The multi-channel EEG signals are low-frequency (low-bandwidth),ranging from $0.1$ Hz to about $100$ Hz, signals. For the purposes of brain activity analysis, EEG signals are typically divided into different frequency bands, namely, $\delta$, $\theta$, $\alpha$, $\sigma$, $\beta$, and $\gamma$. 
These well researched frequency bands have been found to exhibit certain characteristics as shown in Table \ref{tab:eeg-characteristics}, for example an alert and active brain effects the $\beta$ band. These frequency bands are essential in EEG signal processing to understand different brain states and activities based on the patterns observed in these frequency ranges. Analyzing and processing multi-channel EEG data often requires specialized knowledge and training in signal processing, and neuroscience.

 It is well known, from basics of signal processing literature that  signals need to be sampled at twice the maximum frequency present in the signal (also called the Nyquist rate \cite{nyquist1928certain}) to preserve the signal characteristics. Since the frequency bands of interest ($0.1$ to $100$ Hz) are low, most EEG instruments sample the brain activities at rates, typically in the range of $250$ to $1000$ Hz. Observe, in contrast, music, speech and audio signals are sampled at $44.1$ kHz for CD quality and typical sampling rates used in speech and audio signal processing is either $8$ or $16$ kHz.
Formally, the Nyquist rate is a fundamental concept in signal processing, which refers to the minimum rate at which a signal must be sampled to accurately represent the signal without losing any information in terms of fidelity. Specifically, the Nyquist rate is twice the highest frequency (bandwidth) of the signal \cite{nyquist1928certain}.
\begin{table}[!htb]
    \centering
    \scalebox{1}{
    \begin{tabular}{c|c|l}\hline
      Band   &  Frequency (Hz) & Characteristics\\ \hline
       $\delta$  & 0.5–4 & deep sleep and synchronized brain activity \\
       $\theta$ &4–8 & drowsiness, relaxation, and the early stages of sleep\\
$\alpha$ &8-12 & eyes are closed, relaxation and calmness\\
$\sigma$ & 12-16 & associated with sleep spindles during stage 2 sleep \\
$\beta$ &12–30 & active thinking, focus, and alertness\\ 
$\gamma$ &30–100 & cognitive processing, memory, and perception \\ \hline
    \end{tabular}
    }
    \caption{EEG bands, frequency range, and their characteristics.}
    \label{tab:eeg-characteristics}
\end{table}
Our motivation to explore a method for transforming a multi-channel low-bandwidth signal into a single-channel high-bandwidth signal comes from two key observations:
\begin{enumerate}
\item Lack of Pre-trained Models for Low-bandwidth multi-channel EEG: There are currently no large pre-trained models specifically designed for low-bandwidth EEG signals. This is largely due to the lack of standardization in multi-channel EEG data, which varies both in terms of the number of electrodes used and the sampling rates.
\item Availability of Pre-trained Models for High-bandwidth single-channel Signal: In contrast, there are several well-established pre-trained foundational models, such as VGGish, that are available for high-bandwidth single-channel audio signals. These models can be leveraged for various applications, making them highly useful.
\end{enumerate}
By developing a mechanism to transform a low-bandwidth multi-channel EEG signal into a single-channel high-bandwidth signal, we aim to bridge this gap and potentially enable the use of existing pre-trained models for EEG analysis.
Focusing on the question, {\em "Can we use existing pre-trained single-channel (audio or speech) models to analyze or process multi-channel (EEG) signals?"},  in this paper, we propose a transformation, based on simple signal processing, that allows representing a multi-channel low-bandwidth EEG signal as a single-channel high-bandwidth signal. The proposed transformation is bi-directional (reversible) meaning that the single-channel high-bandwidth signal can be converted back into the original multi-channel low-bandwidth EEG signals without any loss of information. 
This is the main contribution of the paper.
The rest of the paper is organized as follows. We describe the proposed transformation to convert a multi-channel low-bandwidth signal into a single-channel high-bandwidth signal in Section \ref{sec:eeg2wav}, we show in Section \ref{sec:experiments} the utility of such a transformation by conducting a set of experiments on a publicly available EEG dataset for odour and subject classification. We conclude in Section \ref{sec:conclusions} and provide future directions.

\section{Transforming a multi-channel signal to a single-channel signal}
\label{sec:eeg2wav}

Consider a  $\nchannel$-channel EEG signal (multi-channel, $\nchannel$ electrodes), of duration $T$ (in seconds) sampled at a sampling frequency of $\efs$ Hz. Let 
$\left\{\eeg_i[\t=1, \cdots, N]\right\}_{i=1}^{\nchannel}$  
denote the $\nchannel$-channel EEG signal 
where $\eeg_i[\t]$ denotes the $n^{th}$ sample in the $i^{th}$ channel 
and $\nofsamples = T\times\efs$. 
The maximum frequency component in $\eeg_i[\t]$ is $(\efs/2)$ which is also the 
bandwidth occupied by each of the individual $\nchannel$-channel's. 
Note that the cumulative bandwidth of the $\nchannel$-channel EEG signal, if one were to stack the $\nchannel$-channel's one on top of other is $(\nchannel \times \efs/2)$, though each individual channel has a bandwidth of $(\efs/2)$. 
The essential idea of the proposed signal transformation is to {\em transform} the  $\nchannel$-channel EEG signals into a single-channel 
signal while retaining the individual characteristics of all the $\nchannel$ channels of the $\nchannel$-channel EEG signal.

Consider $\speech[\tspeech]$ to be the transformed signal, sampled at a
sampling frequency of $\sfs$, namely, the bandwidth of $\speech[\tspeech]$ is $(\sfs/2)$. We now describe a process to transform 
$\{\eeg_i[\t:1,\cdots,N]\}_{i=1}^{\nchannel}$, the $\nchannel$-channel EEG signal, each channel having a bandwidth of $(\efs/2)$ into a single-channel $\speech[\tspeech]$ having a bandwidth of $(\sfs/2)$, namely,
\begin{equation}
\left \{\eeg_i \right\}_{i=1}^{\nchannel} \Longleftrightarrow \speech.
\end{equation}
Note that the transformation is bi-directional, namely, we are able to reconstruct  $\left \{\eeg_i \right\}_{i=1}^{\nchannel}$ from $\speech$ when there is a priori knowledge of how $\speech$ was obtained from $\left \{\eeg_i \right\}_{i=1}^{\nchannel}$. The complete implementation of constructing  $\seeg$ from $ \{\eeg_i\}_{i=1}^{\nchannel}$ is shown in Algorithm \ref{algo:eeg2audio}. We now describe, in detail, the transformation process based on  
simple signal processing. There are essentially four steps in the transformation process.

\begin{figure}[!htb]
    \centering
    \includegraphics[width=0.75\linewidth]{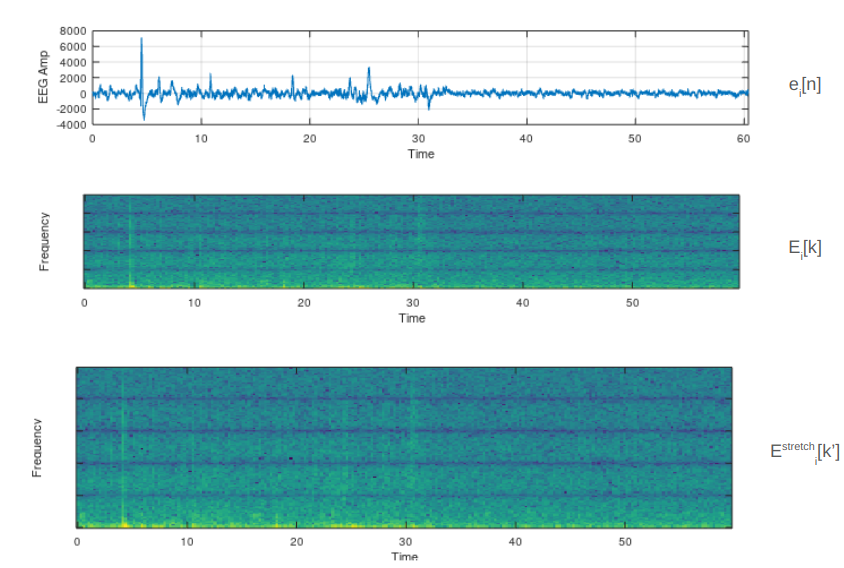}
    \caption{Sample of a single-channel EEG signal of bandwidth $(\efs/2)$ (Top, $\eeg_i$),  spectral representation of $\eeg_i$ (Middle, $\feeg_i$), and  stretched spectral representation with bandwidth $(\sfs/2p)$ (Bottom, $\sfeeg_i$).}
    \label{fig:eeg-ft-stretch}
\end{figure}

\begin{figure}[!htb]
    \centering
    \includegraphics[width=0.75\linewidth]{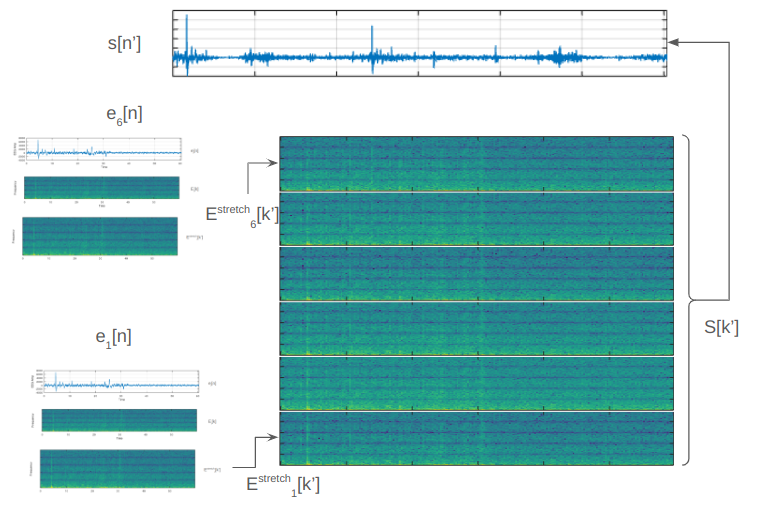}
    \caption{$\nchannel=6$ channel EEG with each channel having a bandwidth of $(\efs/2$) is stretched to $\sfeeg_i$ having a bandwidth $((\efs/2)*6$) and  stacked into a single-channel signal ($\seeg$) with bandwidth $(\sfs/2)$. Implementation details in Algorithm \ref{algo:eeg2audio}.}
    \label{fig:eegtospeech}
\end{figure}

\begin{description}

    \item[Step \#1] 
For each of the $\nchannel$ channels, we compute the well known Fast Fourier Transform (FFT) as shown in Line \ref{algo:fft}, Algorithm \ref{algo:eeg2audio},  namely, 
\begin{equation}
\eeg_i[\t] 
\stackrel{FFT}{\longrightarrow} 
\feeg_i[\f] \;\;\; \text{where, }\;\;\;
\feeg_i[\f] = \sum_{n=0}^{N-1} \eeg_i[\t] \exp^{-j \frac{2\pi}{N} \f \t},
\label{eq:1}
\end{equation}
for $\f = 0, \cdots, (N-1)$ and has the same number of samples as $\eeg_i$. Note that the maximum frequency component in $\feeg_i$ is $(\efs/2)$ and  the $\f^{th}$ sample, $\feeg_i[\f]$ corresponds to the amplitude of the frequency component $\left (\f \times \left (\frac{\efs/2}{(N-1)} \right )\right)$ Hz in the signal $\eeg_i$.
%
%
%

\item[Step \#2] Now we \textit{stretch} $\feeg_i$ so that the stretched $\sfeeg_i$ has a maximum frequency of $({F_s}/{2p})$. This is essentially an index stretching operation, where we take $\feeg_i[\f]$ where $\f$ runs from $1, \cdots, N$ and map those indices to a new set of indices, namely, $\F$ which runs from $1, \cdots, \alpha N$, where $\alpha = ({F_s}/{2p})$. As captured in Line \ref{algo:stretch}, Algorithm \ref{algo:eeg2audio}, the frequency domain signal $\feeg_i[\f]$ is re-indexed to construct $\sfeeg_i[\F]$, namely,
\begin{equation}
\feeg_i[\f] 
\stackrel{\F=\alpha \f} {\longrightarrow} 
\sfeeg_i[\F] 
\label{eq:2}
\end{equation}
where $\alpha = ({F_s}/{2p})$ and $\f=0, \cdots, (N-1)$. Note that the $N$ samples in $\feeg_i$ would result in $\alpha N$ samples in $\sfeeg_i$, namely,
$\F = 0, \cdots, \alpha (N-1)$.  Observe that  the value of $\sfeeg_i[\F]$ is the same as the value of the $\f^{th}$ sample, $\feeg_i[\f]$. However, $\sfeeg_i[\F]$ corresponds to the the frequency component $\left (\alpha \times \f \times \left (\frac{\efs/2}{(N-1)} \right )\right)$ Hz while $\feeg_i[\f]$ corresponds to the frequency component $\left (\f \times \left (\frac{\efs/2}{(N-1)} \right )\right)$ Hz.
While 
the maximum frequency component in $\feeg_i[\f]$ 
is $(\efs/2)$,
the maximum frequency component 
in $\sfeeg_i$ is 
$({F_s}/{2p})$ 
%
This makes sure that when all the $\nchannel$-channels are stacked together, the combined bandwidth of the $\nchannel$-channel's is $({F_s}/{2})$. The process of stretching or reindexing from the implementation perspective is best understood with the help of Algorithm \ref{algo:eeg2audio}. 


\item[Step \#3] We now concatenate or stack the $\nchannel$-channels of
$\{\sfeeg_i[\F]\}_{i=1}^{\nchannel}$ on top of one another as shown in Figure \ref{fig:eegtospeech} to form a single-channel signal of bandwidth $({F_s}/{2})$. Namely,
\begin{equation}
\{\sfeeg_i[\F]\}_{i=1}^{\nchannel} \stackrel{\text{stack}}{\longrightarrow} \fseeg[\F].
\label{eq:3}
\end{equation}
Note that we could stack the $\nchannel$-channels in $p!$ different ways. However, in our implementation, without loss of generality, we choose to stack the $\nchannel$-channels such that the $(i+1)^{th}$ channel was stacked on top of the $i^{th}$ channel as shown in Figure \ref{fig:eegtospeech}. Line \ref{algo:stack_begin} to \ref{algo:stack_end} in Algorithm \ref{algo:eeg2audio} captures the implementation of the stacking  $\{\sfeeg_i[\F]\}_{i=1}^{\nchannel}$ to construct $\fseeg[\F]$.

\item[Step \#4] As the last step in the signal transformation process we compute the inverse FFT (IFFT) of $\fseeg[\F]$ to construct the time domain signal $\seeg[\tspeech]$ (Line \ref{algo:ifft}, Algorithm \ref{algo:eeg2audio}), namely, 
\begin{equation}
\fseeg[\F] \stackrel{IFFT}{\longrightarrow}
\seeg[\tspeech] \;\;\; \text{where, }\;\;\;
\seeg[\tspeech] = \frac{1}{\nofsamples}\sum_{\f'=1}^{\nofsamples} \fseeg[\f'] \exp^{j \frac{2\pi}{N}  \f' \tspeech}.
\label{eq:4}
\end{equation}
Note that the bandwidth of the time domain single-channel signal $\seeg[\tspeech]$ is  $({F_s}/{2})$. 
%
\end{description}
Figure \ref{fig:eeg-ft-stretch} shows the $i^{th}$ channel of the EEG signal $\eeg_i[\t]$ (top), and its spectral representation $\feeg_i[\f]$ (middle) and 
the stretched version of $\feeg_i[\f]$, namely, $\sfeeg_i[\f']$ (bottom). For visual clarity, we use spectrogram of the signals $\eeg_i[\t]$ instead of showing the frequency domain signal $\feeg_i[\f]$. 
Figure \ref{fig:eegtospeech} shows the complete process of constructing a higher bandwidth single-channel signal from low bandwidth multi-channel signals, 
namely,
\begin{equation}
\overbrace{\{\eeg_i[\t]\}_{i=1}^{\nchannel}
\stackrel{FFT}{\longrightarrow}}^{\text{Step \#1}} 
\underbrace{\{\feeg_i[\f]\}_{i=1}^{\nchannel}
\stackrel{\F=\alpha \f} {\longrightarrow}}_{\text{Step \#2}} 
\overbrace{\{\sfeeg_i[\F] \}_{i=1}^{\nchannel}  
\stackrel{\text{stack}}{\longrightarrow}}^{\text{Step \#3}} 
\underbrace{\fseeg[\F]
\stackrel{IFFT}{\longrightarrow}
\seeg[\tspeech]}_{\text{Step \#4}}.
\label{eq:eeg2speech}
\end{equation}

The implementation of the process of transforming a multi-channel signal into a single-channel signal (\ref{eq:eeg2speech}) is captured in Algorithm \ref{algo:eeg2audio}. As mentioned earlier, the transformation process uses fairly well known signal processing techniques and is is self-explanatory. However, we would like to mention that $\text{\tt linspace}(0,\efs,\T')$ (Line \ref{algo:linspace}) is a  well known {\tt NumPy} function that generates $\T'$ evenly spaced values between $0$ and $\efs$. The whole process of stacking or reindexing (\ref{eq:3}) is implemented as shown from Line \ref{algo:stack_begin} to \ref{algo:stack_end} which allows to transfer the amplitude associated with the stretched $i^{th}$ (Line \ref{algo:stretch}) channel frequency of the EEG signal to the \textit{closest} frequency available in the desired signal, $\fseeg$ obtained using (Line \ref{algo:linspace}) described earlier.
It should be noted that by reversing the sequence of operations seen in (\ref{eq:eeg2speech}), we can get back 
    $\{\eeg_i[\t]\}_{i=1}^{\nchannel}$ from  $\seeg[\tspeech]$. Namely, 
    \begin{equation}
    \scalebox{1}{
$\seeg[\tspeech] \stackrel{FFT}{\longrightarrow} 
\underbrace{\fseeg[\F]}_{(\sfs/2)} \stackrel{\text{unstack}}{\longrightarrow} 
\{\underbrace{\feeg_i[\F]}_{(\sfs/2p)}\}_{i=1}^{\nchannel}   
\stackrel{\f=(1/\alpha) \F} {\longrightarrow} \{\underbrace{\feeg_i[\f]}_{(\efs/2)}\}_{i=1}^{\nchannel} 
\stackrel{IFFT}{\longrightarrow} \{\eeg_i[\t]\}_{i=1}^{\nchannel} $}
    \label{eq:speech2eeg}
    \end{equation}
%
%
    Typical values of $\efs$ is around $250$ Hz (EEG signal) while typical $\sfs$ is $8$ or $16$ kHz (audio).
We hypothesize that $\seeg[\tspeech]$ constructed as described in this section represents all the information in the $\nchannel$ channel's, namely, $\{\eeg_i[\t]\}_{i=1}^{\nchannel}$ in a single-channel signal, namely, $\seeg[\tspeech]$. The process of this multi-channel to single-channel transformation allows one to (a) process a single-channel signal $\seeg[\tspeech]$ instead of processing all the $\nchannel$-channels $\{\eeg_i[\t]\}_{i=1}^{\nchannel}$, (b) make use of 
%
any of the known audio processing tools and algorithms, especially when $\sfs$ is chosen to be $8$ or $16$ kHz, and more importantly (c) 
allows the use of existing pre-trained models trained on single-channel signal. 

Fortunately, unlike multi-channel signals, there are quite a few pre-trained models for single-channel signals. For example,  VGGish \cite{hershey2017cnn} is a popular pre-trained model that has been used for audio classification tasks,  YAMNet \cite{yamnet} is another pre-trained deep learning model for sound event detection, DeepSpeech \cite{hannun2014deep} and Wav2Vec \cite{schneider2019wav2vec} are popular pre-trained models used for speech recognition tasks.  While almost all of the pre-trained models have not been trained on EEG \textit{like} data, we reckon that they can be used to extract embeddings or features from the transformed signal $\seeg$ which can be treated as  features corresponding to all the $\nchannel$ channels of EEG, namely, 
$\{\eeg_i\}_{i=1}^{\nchannel}$. 


\begin{algorithm}[!htb]
    \caption{Constructing $\seeg[\tspeech]$ from $ \{\eeg_i[\t]\}_{i=1}^{\nchannel}$}
\begin{algorithmic}[1]
    \Procedure{multi2single}{$\{\eeg_i[1:\T]\}_{i=1}^{\nchannel}, \efs, \sfs$}
    
    \Comment{$\efs:$ sampling rate of $\eeg_i$, $\T:$ number of samples}
    
    \Comment{$\sfs:$ sampling rate of desired signal $\seeg$}
    \State $t_{sec} = \T/\efs$
    \Comment{$t_{sec}:$ duration of $\eeg_i$ in seconds}
    \State $f_{band} = \sfs/2\nchannel$;
    \Comment{$f_{band}:$ bandwidth stretched $e_i$}
    \State $freq[1:\T] = [0:\T-1]\times \frac{\efs}{(\T-1)}$

    \Comment{freq corresponding to $\T$ sample}
    \State $\T' = t_{sec}* \sfs$;
    \Comment{Samples in the single-channel} 
 \State $F_{desired}[1:\T'] = \text{\tt linspace}(0,\sfs,\T')$; \label{algo:linspace}
\State $\fseeg[1:\T'] = 0$; $\seeg[1:\T']=0$\Comment{Initialize}
    \For{$i=1:\nchannel$}
        \State $\feeg_i[1:\T] = \text{FFT}(\eeg_i[1:\T])$ \label{algo:fft}
        \Comment{Eq (\ref{eq:1}); Step \#1} 
        \State $l_{f} = (i-1)* f_{band}$; 
        \State $f_{stretch}[1:\T] = l_{f} + \left (freq[1:\T]\times\frac{f_{band}}{\efs} \right )$ \label{algo:stretch}

        \Comment{Eq (\ref{eq:2}); Stretching.}
        \For{$j=1:\T$} \label{algo:stack_begin}
        \Comment{Eq (\ref{eq:3}); Stacking}
        \State $min_v = 10^{10}; min_{1} = -1; min_{2}=-1$
         \For{$k=1:\T'$}
         \If{$(|F_{desired}[k] -  f_{stretch}[j]| \le min_v)$}
        \State $min_v \leftarrow |F_{desired}[k] -  f_{stretch}[j] |$
        \State $min_{1} \leftarrow j; min_{2} \leftarrow k$ 
        \EndIf
         \EndFor
         \State $\fseeg[min_{2}] = \feeg_i[min_{1}]$;
             \EndFor \label{algo:stack_end}
         
      \EndFor
      \State $\seeg[1:\T'] = \text{IFFT}(\fseeg[1:\T'])$  \label{algo:ifft}
      \Comment{Eq (\ref{eq:4}); Construct $\seeg$ from $\fseeg$}
    \EndProcedure
\end{algorithmic}
\label{algo:eeg2audio}
\end{algorithm}

\section{Experimental Analysis}
\label{sec:experiments}
We use the widely used, publicly available odour-\EEG\ dataset (\ieeedata) \cite{59nx-6g46-22} for training (a) odour-classification \cite{DBLP:conf/embc/PandharipandeTC23} system and (b) subject identification \cite{10290059} system. The \ieeedata\  dataset consists of a $32$ channel \EEG\ signal recorded using the Cerebus system from $11$ healthy individuals ($8$ males and $3$ females), right-handed, aged $24.9 \pm 3.0$ years in response to $13$ odour stimuli (rose, caramel, rotten smell, canned peaches, excrement, mint, tea, coffee, rosemary, jasmine, lemon, vanilla, and lavender). 
Two of  the $32$ channels are reference channels, making it, $\nchannel=30$ usable channels for analysis. The electrodes for collecting \EEG\ are arranged according to the international $10$-$20$ system, and sampled at $\efs =1$ kHz. Each sample  
collected was for a duration of $10$ seconds ($N = 10000$ samples), called a trial. 
In total, \ieeedata\ dataset has $11$ (\participants) $\times$ $13$ (odours) $\times$ $35$ (trials) resulting in a total of $5005$ $30$-channel \EEG\ samples. We partition the \ieeedata\ dataset into non-intersecting $80:20$ train and test sets. The training set consists of $4004$ samples while the remaining $1001$  samples were used for test. We retained the split in all our experiments for consistency in experiments.


 We converted the $30$-channel ($\nchannel = 30$) \EEG\ signal into a single-channel high bandwidth audio signal of $\sfs = 16$ kHz as described in the previous section following Steps \#1, \#2, \#3, and \#4 implemented as mentioned in Algorithm \ref{algo:eeg2audio}. 
In the first set of experiments, using the single-channel high bandwidth ($16$ kHz) transformed signal, we extracted the spectrogram (for example, Figure \ref{fig:eeg-ft-stretch} (middle) is the spectrogram of the $\eeg_i$)
using $64$ msec ($1024$ samples) as the window and $48$ msec ($768$ samples) as the overlap. 
%
The resulting size of the spectrogram was $513 \times 621$, which is a representation of all the $30$ \EEG\ channels {\em simultaneously}. So each \EEG\ trial was represented by a spectrogram matrix of size $513 \times 621$. We used a Convolutional Neural Network (CNN) architecture \cite{lecun1998gradient}
as shown in Table \ref{tab:cnn_model} for the purpose of odour ($13$ class, 157945997 trainable parameters) and subject ($11$ class, 157945739 trainable parameters) classification. The CNN was implemented in {\tt keras} and no hyper parameter tuning was done. As is common, $20\%$ of the train data was used for the purposes of validation. So we had $3203$, $801$, $1001$ samples for training, validation and test respectively with no overlap.

\begin{table}[!htb] 
\centering

%
{
\centering
\begin{tabular}{lll} \hline
 Layer (type)                  &   Output Shape              &     Param \#     \\ \hline 
 spectrogram (Conv2D)          &   (None, 511, 619, 32)      &     320         \\ 
 max\_pooling2d (MaxPooling2D)  &   (None, 255, 309, 32)      &     0           \\ 
 conv2d (Conv2D)               &   (None, 253, 307, 64)      &     18496       \\ 
 max\_pooling2d (MaxPooling2D)&   (None, 126, 153, 64)      &     0           \\ 
 flatten (Flatten)             &   (None, 1233792)           &     0           \\ 
 dense (Dense)                 &   (None, 128)               &     157925504   \\ 
 {\textit{[odour]}} {subject} (Dense)               &   (None, {\textit{[13]}} 11)                &     {\textit{[1677]}} 1419        \\ \hline
\end{tabular}
}

\caption{2D-CNN architecture used 
for {[Odour]} and Subject Classification. Using spectrogram features extracted from the transformed single-channel signal $\speech$.}
\label{tab:cnn_model} 
\end{table}

\begin{figure}[!htb]
    \centering
    \begin{subfigure}{0.49\textwidth}
    \includegraphics[width=\textwidth]{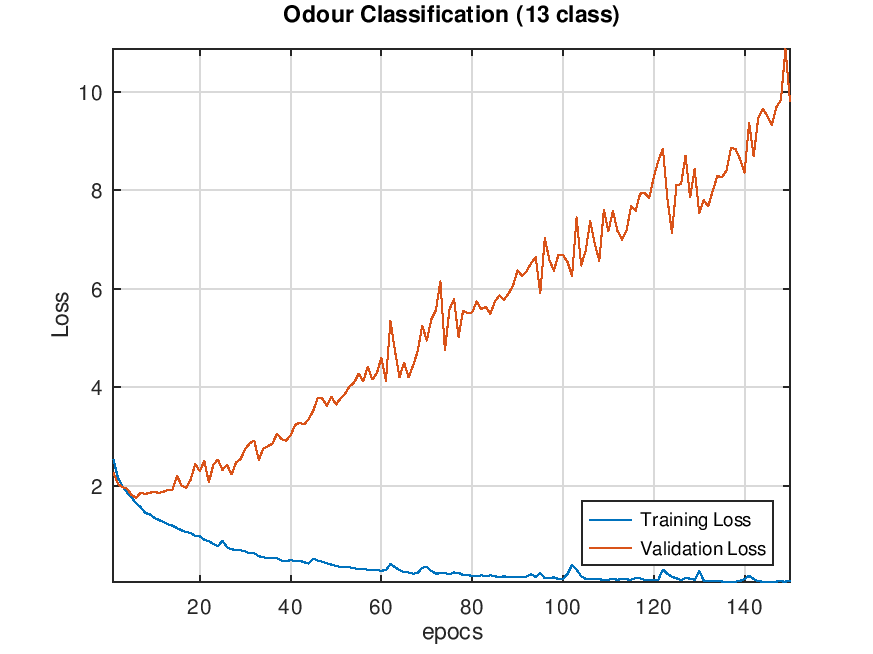}
        \caption{Odour. Loss.}
    \label{fig:oc_loss}
\end{subfigure}
\hfill
    \begin{subfigure}{0.49\textwidth}
    \includegraphics[width=\textwidth]{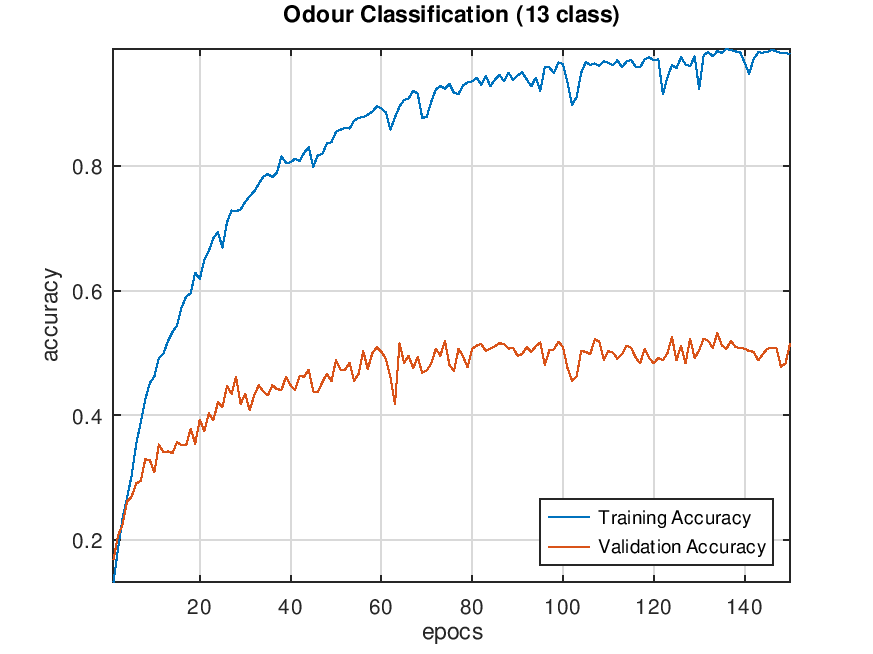}
        \caption{Odour. Accuracy.}
    \label{fig:oc_acc}
\end{subfigure}
    \begin{subfigure}{0.49\textwidth}
    \includegraphics[width=\textwidth]{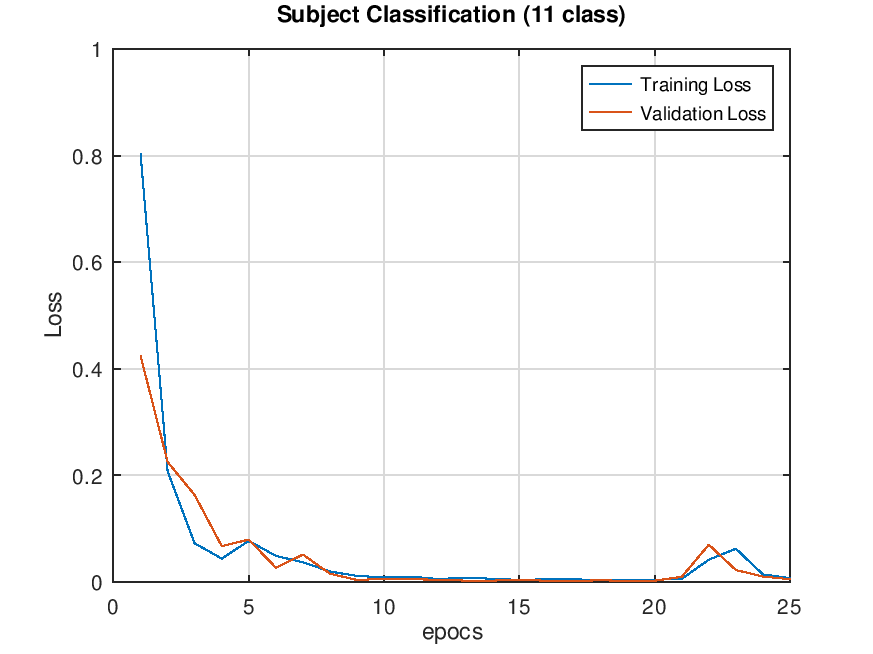}
    \caption{Subject. Loss.}
    \label{fig:sc_loss}
\end{subfigure}
\hfill
    \begin{subfigure}{0.49\textwidth}
    \includegraphics[width=\textwidth]{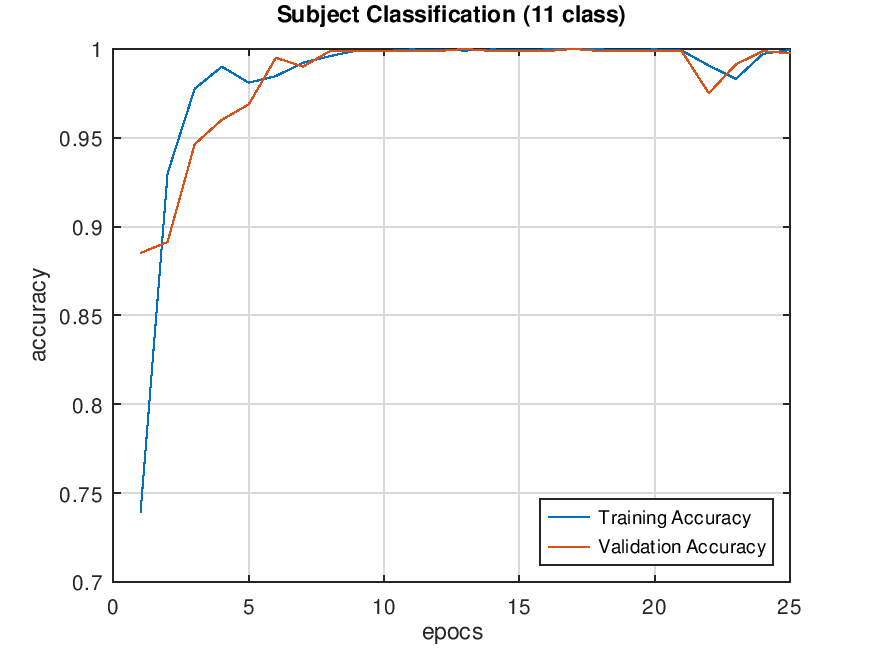}
            \caption{Subject. Accuracy.}
    \label{fig:sc_acc}
\end{subfigure}

    \caption{Odour, Subject Classification. Train, Validation plots. For 2D-CNN architecture (Table \ref{tab:cnn_model}) using spectrogram of the transformed signal as the input features.}
    \label{fig:oc_loss_acc}
\end{figure}

 The CNN architecture (Table \ref{tab:cnn_model}) with a batch size of $32$ was run for $150$ epochs for odour classification.  The training and validation losses (Figure \ref{fig:oc_loss}) and accuracy is shown in Figure \ref{fig:oc_acc}. The test accuracy for $13$ class odour classification was  $51.85\%$. 
The CNN architecture (Table \ref{tab:cnn_model}) with a batch size of $32$ was run for $25$ epochs for subject ($11$ class) classification.  The training and validation losses are shown in Figure \ref{fig:sc_loss} while the accuracy is shown in Figure \ref{fig:sc_acc} for subject classification. The test accuracy for $11$ class subject classification was  $99.70\%$. As seen in  Table \ref{tab:performance}, we achieve the best performance for both odour and subject classification when a single-channel $\speech$ is used instead of the multi-channel $\{\eeg_i\}_{i=1}^{30}$. Also seen in Table \ref{tab:performance} are the performance of SVM ($27.63\%$) and RF ($29.85\%$) classifiers on hand crafted features for odour classification as reported in \cite{DBLP:conf/embc/PandharipandeTC23} and subject classification \cite{10290059}. The total number of features used were $720$ which included energy, entropy, discrete wavelet transform for each of the bands.
\begin{table}[!htb]

    \centering
    \scalebox{1}{
    \begin{tabular}{|c|c|c|c|c|}\hline
     Signal & Features (size)   &  Classifier & Odour & Subject \\ \hline
     $\seeg$& Spectrogram ($513 \times 621$) & 2D-CNN &  {\bf 51.85} &  {\bf 99.70} \\ \hline 
     
     $ \{\eeg_i\}_{i=1}^{30}$ &  \EEG\ features ($720$) & RF & 29.85 \cite{DBLP:conf/embc/PandharipandeTC23} &  97.94 \cite{10290059} \\ 
     $ \{\eeg_i\}_{i=1}^{30}$ & \EEG\ features ($720$) & SVM & 27.63 \cite{DBLP:conf/embc/PandharipandeTC23}& 93.55 \cite{10290059} \\ 
     \hline \hline
    \end{tabular}
    }
    \caption{Odour and Subject 
    Classification when the transformed single-channel ($\seeg$) is used versus multi-channel signal ($ \{\eeg_i\}_{i=1}^{30}$) for different features and classifiers.}
    \label{tab:performance}
\end{table}


As mentioned earlier, one of the benefits of transforming a multi-channel \EEG\ signal into a single-channel high bandwidth signal is that it 
allows the use of existing pre-trained models trained on massive amounts of single-channel data. The use of pre-trained models allows to extract  embeddings from these models which can be used as features rather than using handcrafted features for the downstream classification tasks. 
In the next set of experiments, we used {\tt VGGish} \cite{vggish_model_ckpt} and {\tt YAMNet} \cite{yamnet_model} pre-trained model  
to extract embeddings, because both of these models have been trained for acoustic or audio event detection. We discounted the use of other pre-trained models like {\tt Wav2Vec} or {\tt DeepSpeech} because they are specifically trained for human speech; we feel that the transformed \EEG\ signal is closer to an acoustic event like signal compared to human speech. 

\begin{figure}[!htb]
    \centering
    \begin{subfigure}{0.49\textwidth}
    \includegraphics[width=\textwidth]{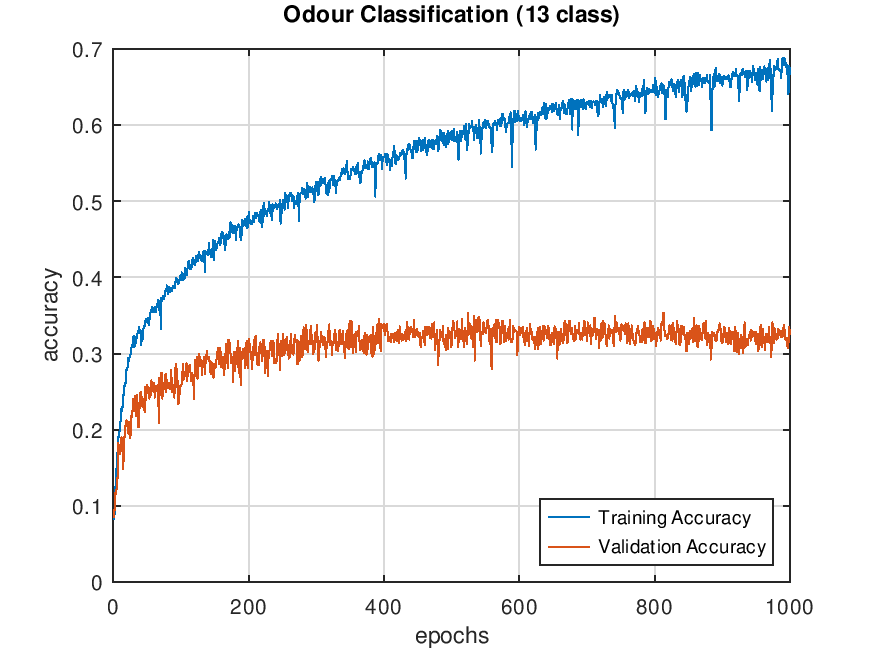}
        \caption{Odour Classification.}
    \label{fig:vggish_oc_acc}
\end{subfigure}
\hfill
    \begin{subfigure}{0.49\textwidth}
    \includegraphics[width=\textwidth]{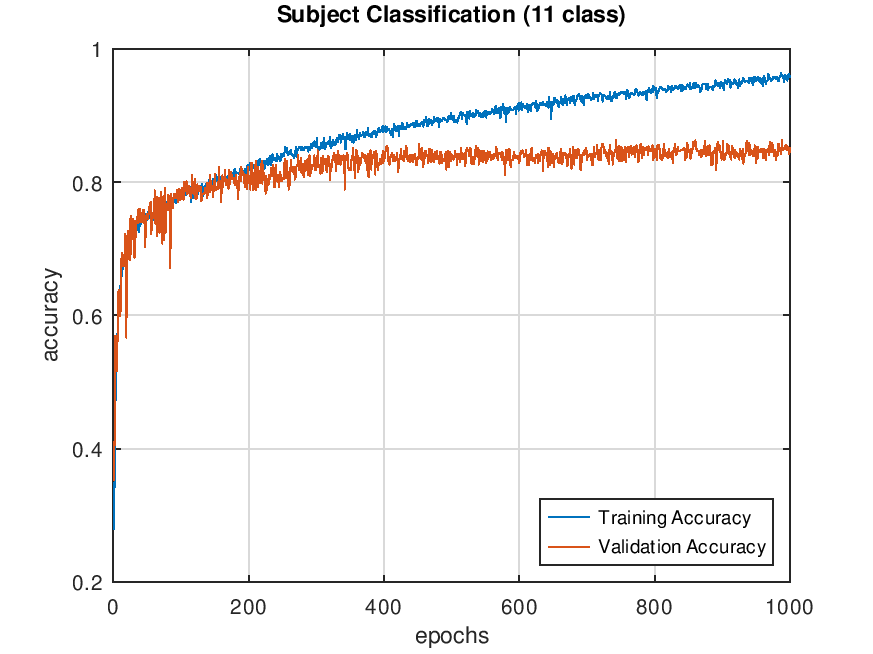}
        \caption{Subject Classification.}
    \label{fig:vggish_sc_acc}
\end{subfigure}
    \caption{Odour and Subject classification train and validation accuracy plots for 2D-CNN architecture (Table \ref{tab:cnn_model}(a)) using  {\tt VGGish} embeddings.} 
    \label{fig:vggish_acc}
\end{figure}

The {\tt VGGish} model gives an embedding of size $128$ dimension at $1$ Hz. 
As a consequence, the $10$ second $30$-channel \EEG\ signal, transformed to a single-channel signal ($\speech$), had a representation in the form of a feature vector of size $10 \times 128$.  
We used a shallow 2D-CNN architecture as shown in Table \ref{tab:vggish_cnn}(a). As seen in Table \ref{tab:performance_pretrained} the performance of the odour classification ($2066571$ trainable parameters) was $30.27\%$ while the performance was $85.91\%$ for subject classification ($2066829$ trainable parameters). In both the cases, we did not do any hyper-parameter turning and used a batch size of $32$ and trained for $1000$ epochs. The training and validation accuracy are shown in Figure \ref{fig:vggish_oc_acc} and  Figure \ref{fig:vggish_sc_acc} for odour and subject classification respectively. Along similar lines, the {\tt YAMNet} model resulted in an embedding of size $20\times521$, where $521$ corresponds to the number of audio event classes that {\tt YAMNet} was trained for. Using a shallow 2D-CNN as mentioned in Table \ref{tab:vggish_cnn}(b) we trained for both odour ($19097997$ trainable parameters) and subject classification ($19097739$ trainable parameters). As seen in Table \ref{tab:performance_pretrained} the performance of a 2D-CNN architecture using {\tt YAMNet} embeddings derived from the  transformed signal ($\speech$) was $33.17\%$ and $94.21\%$ for odour and subject classification respectively. The performance for both odour and subject classification is better when {\tt YAMNet} pre-trained model is used compared to when {\tt VGGish} is used; the improvement when  {\tt YAMNet} pre-trained model is used shows a relative improvement of $\approx 9\%$ for both  odour 
$\left (
  \left (
    \frac{33.17-30.37}{30.27} 
   \right ) \times 100 
  \right )$ 
and subject 
$\left (
 \left (
  \frac{94.21-85.91}{85.91} 
  \right ) \times100 
 \right )$ classification.
However, as seen in Table \ref{tab:performance_pretrained}, while the performance, when a pre-trained model is used, is not as good as when the spectrogram features ($513 \times 621$) of $\speech$ is used, it nevertheless demonstrates the effective use of existing pre-trained audio models, even though they have not been explicitly trained to process and thereby extracting features low bandwidth multi-channel EEG signals. It should be noted that one of the challenges in processing \EEG\ signals is not only the ability to select as appropriate channel or a set of channels (from the multi-channel signal data) but also the ability to determine the right set of hand-crafted features that best suit the task. Using pre-trained models on the transformed signal $\speech$ eliminates this challenge. 
%
This experimental result demonstrates the effective use of pre-trained models to enable building shallow machine learning models (compare the architecture in Table \ref{tab:cnn_model} and Table \ref{tab:vggish_cnn}) for odour and subject classification.  

\begin{table}[!htb]
    \centering


\begin{tabular}{lll}\\ \hline
 Layer (type)                  &   Output Shape              &     Param \#     \\ \hline
 vggish (Conv2D)               &   (None, 8, 126, 64)        &     640         \\ 
 max\_pooling2d\_2 (MaxPooling2D)&   (None, 4, 63, 64)         &     0           \\ 
 flatten\_1 (Flatten)           &   (None, 16128)             &     0           \\ 
 dense\_1 (Dense)               &   (None, 128)               &     2064512     \\ 
 {[odour]} {subject} (Dense)               &   (None, {[13]} 11)                &     {[1677]} 1419        \\ \hline
\end{tabular}
\centerline{(a) 2D-CNN architecture with {\tt VGGish} embeddings.}

\begin{tabular}{lll}\\ \hline
 Layer (type)                  &   Output Shape              &     Param \#     \\ \hline
 yamnet (Conv2D)            & (None, 18, 519, 64)  &     640 \\       
 max\_pooling2d\_2 (MaxPooling2D)&   (None, 9, 259, 64)     &   0         \\
  flatten\_1 (Flatten)           &           (None, 149184)  &          0   \\      
 dense\_1 (Dense)               &   (None, 128)             &  19095680  \\
 {[odour]} {subject} (Dense)               &      (None,  {[13]} 11) &               {[1677]} 1419 \\      \hline
                                                                %
\end{tabular}
\centerline{(b) 2D-CNN architecture with {\tt YAMNet} embeddings.}

\caption{Shallow 2D-CNN architecture 
using (a) {\tt VGGish} embeddings and (b) {\tt YAMNet} embeddings derived from the transformed single-channel signal $\speech$.}
    \label{tab:vggish_cnn}
\end{table}

For the purpose of comparison, we use the original $30$-channel \EEG\ signal to extract spectrogram for each of the $30$  \EEG\ channels separately using {\tt librosa}'s short term Fourier transform,  to obtain $30$ spectrograms of dimension $62\ \text{freqency bins} \times 141\ \text{time frames}$,
in effect we had a $30 \times 62 \times 141$ tensor representing the \EEG\ signal.
 We used a 3D-CNN (see Table \ref{tab:model-3dcnn}) architecture consisting of $2$, convolution followed by average pooling layers followed by Flatten, Dense, BatchNormaization, Dropout(0.5) and Dense layers to train a model for odour and subject classification.  We used the default hyper-parameters with a batch size of $32$ and $100$ epochs. The performance for odour and subject classification was $26.17\%$ and $96.1\%$ respectively (see Table \ref{tab:performance_spectrogram}). Clearly, deriving spectrogram features from the the transformed single-channel signal ($\speech$) performs better ($99.70\%$) than the same spectrogram features derived from the original multi-channel $\{\eeg_i\}_{i=1}^{30}$ signal ($96.10\%$) for subject classification. Similar enhanced performance is observed for odour classification, compare $51.85\%$ to $26.17\%$ (Table \ref{tab:performance_spectrogram}). This enhanced performance can be attributed to the proposed signal transformation process that allows the deep learning architecture to \textit{view all the $30$-channels together} as a single unit rather than have them viewed as a stack of $30$ spectrograms. 

\begin{table}[!htb]
    \centering
    \scalebox{1}{
    \begin{tabular}{|c|c|c|c|c|}\hline
     Signal & Features (size)   &  Classifier & Odour & Subject \\ \hline
     $\seeg$& Spectrogram ($513 \times 621$) & 2D-CNN &  {\bf 51.85} &  {\bf 99.70} \\ \hline 
     $\seeg$&{\tt VGGish} ($10 \times 128$) & 2D-CNN &  30.37 & 85.91\\ 
     $\seeg$&{\tt YAMNet} ($20 \times 521$) & 2D-CNN &  33.17 & 94.21\\ \hline\hline
     
     $\seeg$&{\tt VGGish} ($1280$) & RF & 38.26 & 87.01\\ 
     $\seeg$&{\tt VGGish} ($1280$) & SVM & 16.98 & 68.53\\ \hline
     
     $\seeg$&{\tt YAMNet} ($10420$) & RF & 38.66 & 93.91\\ 
     $\seeg$&{\tt YAMNet} ($10420$) & SVM & 25.77 & 88.71\\ 
     
     \hline \hline
    \end{tabular}
    }
    \caption{Odour and Subject 
    Classification using the transformed single-channel signal ($\speech$) to demonstrate the use of existing pre-trained single-channel models ({\tt VGGish} and {\tt YAMNet}.}
    \label{tab:performance_pretrained}
\end{table}

\begin{table}[!htb] \begin{tabular}{lll}\\ \hline
Layer (type)                  &   Output Shape              &     Param \#     \\ \hline
conv3d (Conv3D)               &   (None, 28, 60, 139, 32)   &     896         \\ 
average\_pooling3d (AveragePooling3D) &   (None, 14, 30, 69, 32)    &     0           \\ 
conv3d\_1 (Conv3D)             &   (None, 12, 28, 67, 64)    &     55360       \\ 
average\_pooling3d\_1 (AveragePooling3D)&   (None, 6, 14, 33, 64)     &     0           \\ 
conv3d\_2 (Conv3D)             &   (None, 4, 12, 31, 128)    &     221312      \\ 
max\_pooling3d (MaxPooling3D)  &   (None, 2, 6, 15, 128)     &     0           \\ 
flatten (Flatten)             &   (None, 23040)             &     0           \\ 
dense (Dense)                 &   (None, 512)               &     11796992    \\ 
batch\_normalization (BatchNormalization)&   (None, 512)               &     2048        \\ 
dropout (Dropout)             &   (None, 512)               &     0           \\
{[odour]} subject (Dense)               &   (None, {[13]} 11)                &     [6669] 5643       \\ \hline
\end{tabular} 
\caption{3D-CNN Architecture used with $\{\eeg_i\}_{i=1}^{30}$. Total trainable parameters for odour and subject classification is $12083277$ and $12082251$ respectively.}
\label{tab:model-3dcnn}  \end{table}

\begin{table}[!htb]
    \centering
    \scalebox{1}{
    \begin{tabular}{|c|c|c|c|c|}\hline
     Signal & Features (size)   &  Classifier & Odour & Subject \\ \hline
     $\seeg$& Spectrogram ($513 \times 621$) & 2D-CNN &  {\bf 51.85} &  {\bf 99.70} \\
     $ \{\eeg_i\}_{i=1}^{30}$ & Spectrogram ($ 30 \times 62 \times 141 $) & 3D-CNN & 26.17 & 96.10 \\ 
     \hline \hline
    \end{tabular}
    }
    \caption{Odour and Subject 
    Classification using spectrogram derived from the transformed signal $\speech$ and the original multi-channel signal $\{\eeg_i\}_{i=1}^{30}$.}
    \label{tab:performance_spectrogram}
\end{table}

To sum up the experimental observations as seen from Table \ref{tab:performance}, Table \ref{tab:performance_pretrained}, and Table \ref{tab:performance_spectrogram}, 
\begin{enumerate}
    \item The best performance for odour (subject) classification is $51.85\%$   $(99.70\%)$ is obtained using the spectrogram features extracted from the single-channel transformed signal ($\seeg$) using a 2D-CNN architecture. The improved performance compared to any features (spectrogram or handcrafted) extracted from the $30$ individual EEG channels ($\eeg_i$) demonstrates the usefulness of transforming a multi-channel low-bandwidth signal into a single-channel signal.
\item While both {\tt YAMNet} and  {\tt VGGish} are pre-trained models that are used for audio classification tasks, the use of {\tt YAMNet} embeddings show better performance  compared to use of {\tt VGGish} embeddings across different classifiers for both odour and subject classification tasks (Table \ref{tab:performance_pretrained}). While the relative improvement in performance is around $8-9\%$ for 2D-CNN and RF classifiers it is much higher ($34\%$ for subject and $51\%$ for odour classification) for SVM classifier.   
\item The performance of both odour and subject classification  task using embeddings derived from pre-trained models ({\tt VGGish}, {\tt YAMNet}) from the transformed signal ($\seeg$) is lower compared to the using spectrogram as features extracted from either $\speech$ or $\{\eeg_i\}_{i=1}^{30}$. This is to be expected because the spectrograms are in some sense a better representation of the signal compared to the embeddings extracted from pre-trained models which have been trained on audio signals that are different from the EEG signals. 
\item  The best performance using pre-trained models is achieved using {\tt YAMNet} embeddings + 2D-CNN for subject classification ($94.21\%$) and {\tt YAMNet} embeddings + RF for odour classification ($38.66\%$) as seen in Table \ref{tab:performance_pretrained}.
\item Use of {\tt YAMNet} or {\tt VGGish} features demonstrate superior performance (except SVM classifier) compared to extracting hand crafted features from all the $30$ EEG channels (Table \ref{tab:performance} and \ref{tab:performance_pretrained}). This demonstrates that the features extracted as embedding from the single-channel signal ($\seeg$) using models not specifically trained for EEG signals perform better than even the hand-crafted signal derived from the multi-channel signal ($\eeg_i$). 

\item The difference in performance between odour (best $51.85\%$) and subject (best $99.70\%$) classification can be attributed to the \ieeedata\ data.
Fig. \ref{fig:tsne} shows the t-SNE plot \cite{maaten2008visualizing} for the training data;  better clustered for subject (Fig. \ref{fig:sc_tsne}) than for odour (Fig. \ref{fig:oc_tsne}) correlates with the poorer performance for odour classification. 

\begin{figure}[!htb]
    \centering
    \begin{subfigure}{0.49\textwidth}
        \includegraphics[width=\linewidth]{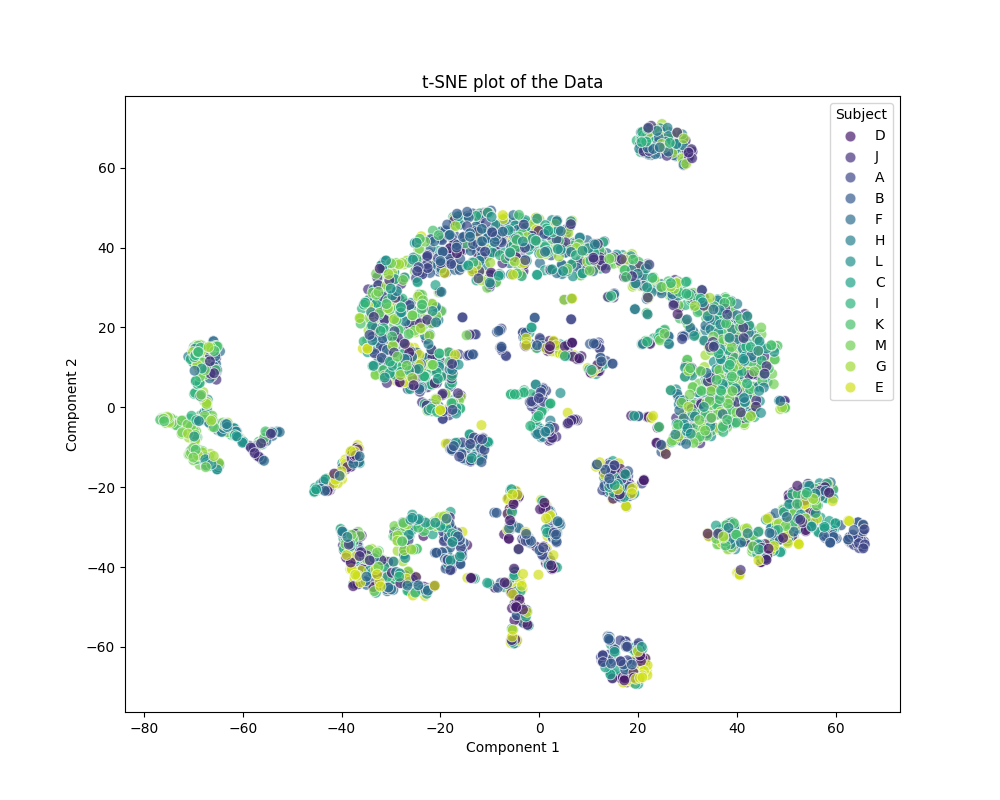}
        \caption{Odour}
        \label{fig:oc_tsne}
    \end{subfigure}
\hfill
    \begin{subfigure}{0.49\textwidth}
        \includegraphics[width=\linewidth]{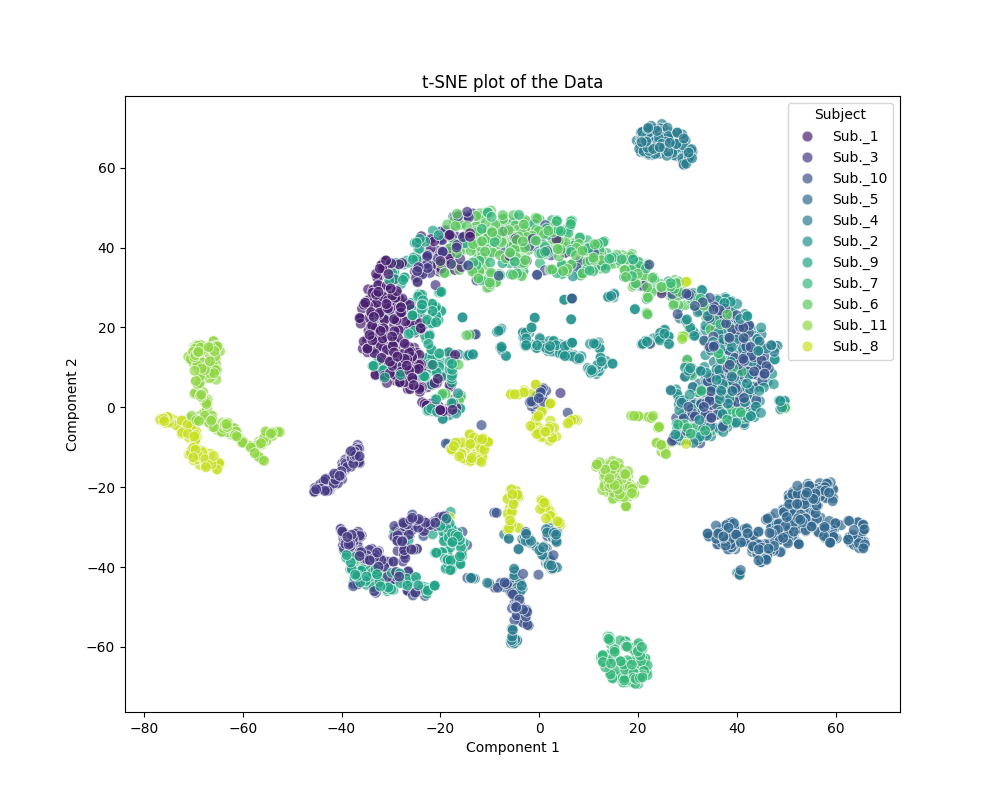}
        \caption{Subject}
        \label{fig:sc_tsne}
    \end{subfigure}
    
    \caption{t-SNE plot of the Training data used for (a) odour classification, (b) for subject classification (better clustering).}
    \label{fig:tsne}
\end{figure}
\end{enumerate}

\section{Conclusions}
\label{sec:conclusions}
In this paper, we proposed a simple yet effective signal processing method to transform a  multi-channel low bandwidth \EEG\ signal into a single-channel high bandwidth signal. The signal transformation is such that it retains all the properties of the original multi-channel signal in the single-channel signal; further the transformation is bi-directional, meaning it allows for reconstructing the multi-channel signal for the single-channel signal.  This is the main contribution of this paper.
We motivated the need for such a transformation, primarily because the signal transformation allows for a comprehensive, time synchronized view of the multi-channel signal in a single-channel waveform. This is useful, especially in an end-to-end deep learning setup, because there is no extra burden on the end-to-end deep learning setup to \textit{learn the association between different channels} in the multi-channel signal. Single-channel representation derived from the multi-channel signal allows for  all the channels to be simultaneously available for processing. Another motivation is that this transformation enables exploit the use of existing pre-trained single-channel acoustic models especially because of the absence of any such pre-trained models for multi-channel signals like EEG.
Experimental results show that the single-channel high bandwidth signal is able to retain the properties of the \EEG\ signal very effectively; indicated by the superior performance for both odour and subject classification compared to hand crafted features extracted from multi-channel \EEG\ signal.

\subsection{Discussion}
While we have experimented with varying number of convolution layers, different convolution sizes, and varying number of nodes in the dense layers and presented the best performing architectures, we have not experimented with hype-parameter tuning. There is plenty of scope for further experimentation using   
different deep learning architectures and hype-parameter tuning. However, as emphasized earlier, the focus of this paper was on demonstrating that a simple signal processing transformation allowed for a better representation of a multi-channel signal. 

The use of pre-trained models trained on acoustic data for acoustic event detection of classification task in the context of \EEG\ data does throw up several questions including the relevance of these pre-trained models for \EEG\ data processing. However, listening to the transformed signal ($\seeg$) derived from ($\eeg_i$) made us believe that the transformed signal might have some relevance to acoustic events. The experimental results seem to demonstrate that the pre-trained models are able to extract embeddings that seem to capture the odour and the subject characteristics in the transformed signal. 
%
As future work, the single-channel data is neither speech nor audio but converting EEG multi-channel data into a single-channel data, as mentioned in this paper, might allow building models which best work for EEG data.

While both the pre-trained models have been trained for downstream acoustic event detection or classification, it is not clear why one works better than the other. While the focus in this paper was to use the pre-trained models as a black box to extract embeddings or features and not to looking at the architecture or training of the pre-trained models, as future work, it might be a good idea to understand and correlate the performance of odour and subject classification to the architecture and training data associated with the pre-trained model.


\bibliographystyle{cas-model2-names}

\bibliography{mybib}



\end{document}